\documentclass[aps,prl,twocolumn,superscriptaddress]{revtex4}

\usepackage{amsmath}
\usepackage{amssymb}
\usepackage{graphicx}

\newcommand{\obs}{\ensuremath{\hat{\cal O}}}

\pdfoutput=1

\newcommand{\eq}{\begin{equation}}
\newcommand{\eeq}{\end{equation}}
\newcommand{\ket}[1]{\left|#1\right\rangle}
\newcommand{\bra}[1]{\left\langle #1\right|}

\graphicspath{{./images/}}

\begin{document}

\newcommand{\FigureOne}{
\begin{figure}
\centering
\includegraphics[width=\columnwidth]{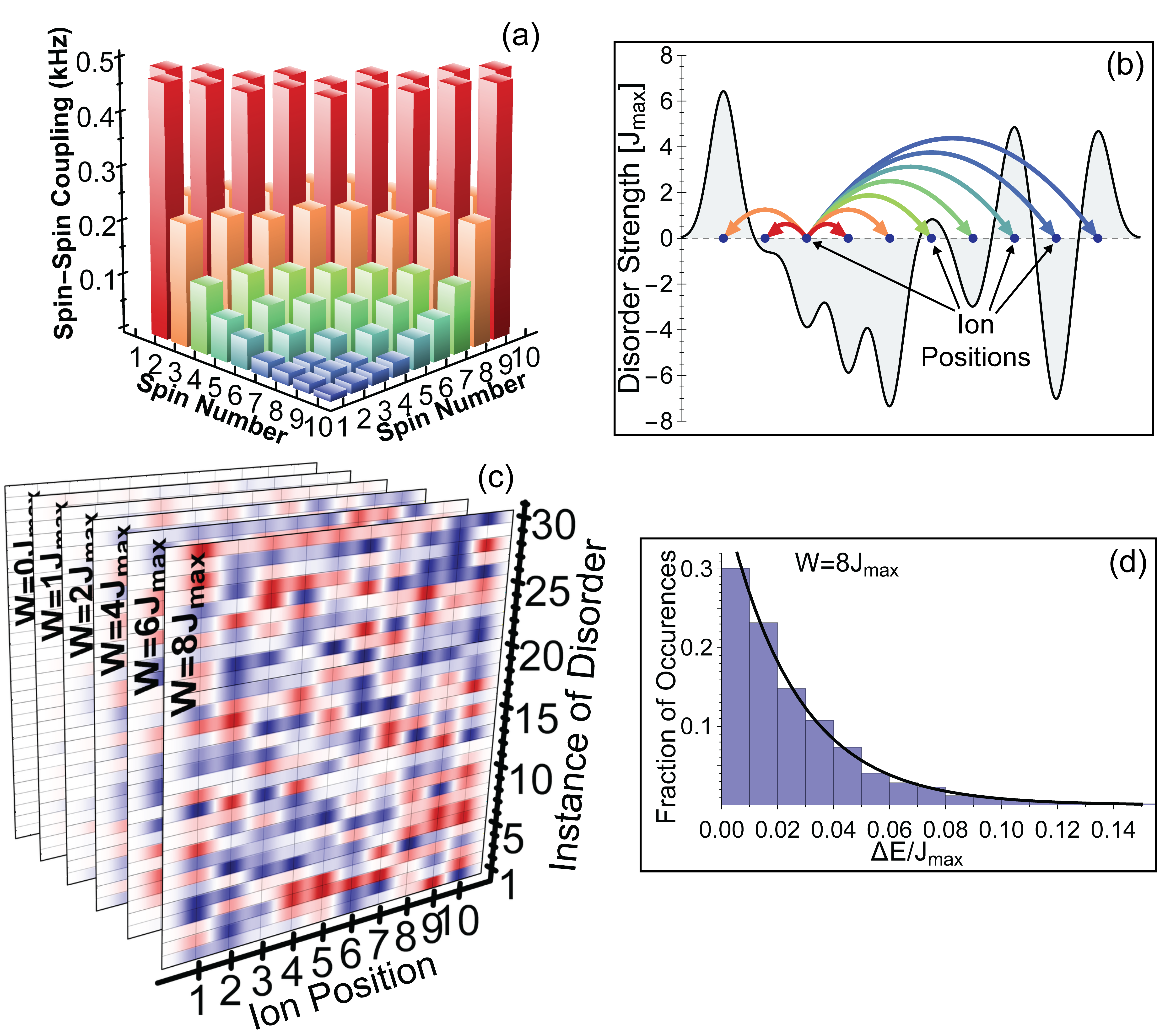}
\caption{\textbf{An Interacting spin model with random disorder} (a) Directly measured elements of the spin-spin coupling matrix $J_{ij}$ (Eq. (1)). The long range interactions decay as $J_{max}/r^{1.13}$.  (b) A specific instance of the random disordered field with a schematic illustration of the long-range interactions and (c) the random values of the disordered field for all 30 instances of disorder for several different disorder strengths and for each ion. (d)  The level statistics calculated from the measured spin-spin coupling matrix (a) and applied disorders (c) are Poisson-distributed (black line is the expected level spacings for a Poisson distribution), as predicted for a MBL system.  }

\label{fig:Interacting}
\end{figure}
}

\newcommand{\FigureTwo}{
\begin{figure*}
\includegraphics[width=18cm]{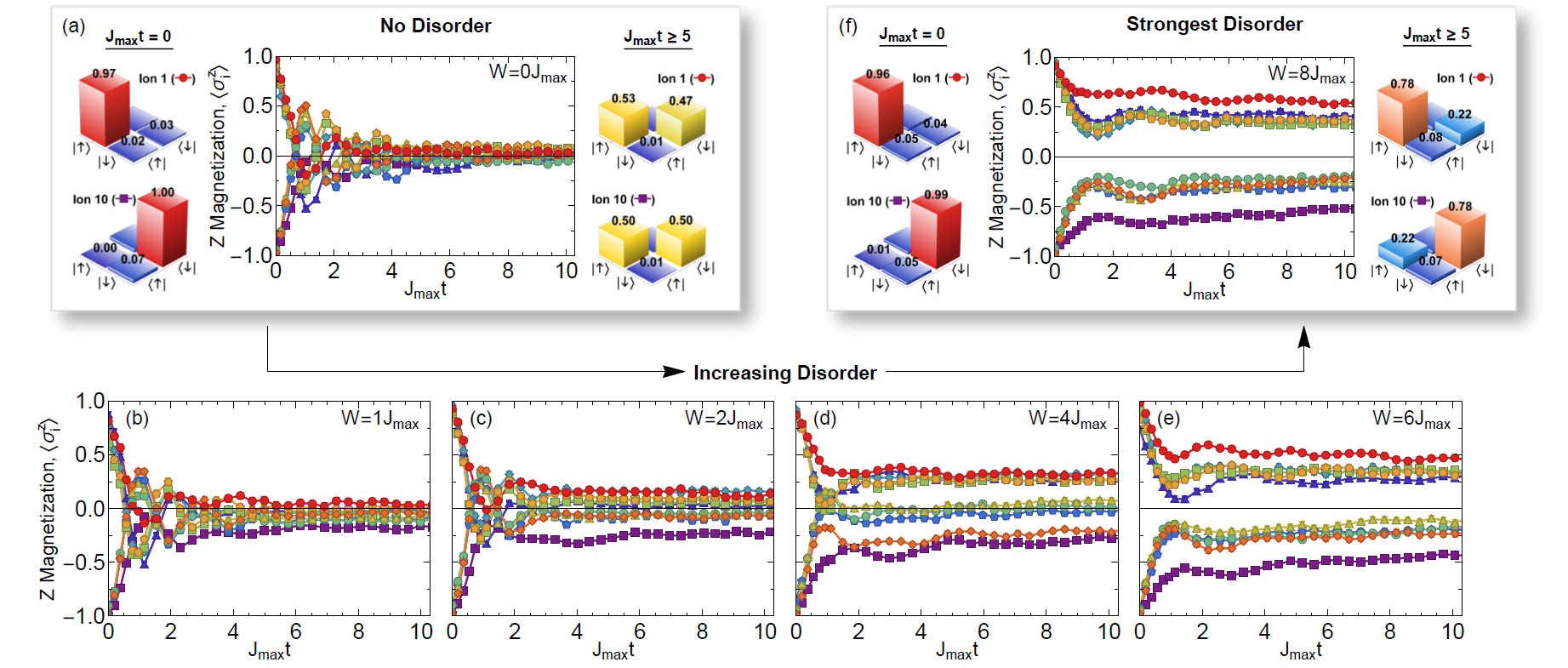}
\caption{\textbf{Emergence of a Many-Body Localized State.} (a) shows the time-evolved single-site magnetizations $\langle\sigma_{i}^{z}\rangle$  (different colors represent different ions) for the Hamiltonian in Eq. (1) and with $B=4J_{max}$ with no applied disorder ($D_{i}=0$). The initial-state reduced density matrices for ions 1 and 10 show the spins start in a product state along the $z$ direction. The time-averaged reduced density matrices for $J_{max}t>5$  agree with the values predicted by the ETH, implying the system has thermalized locally. (b-e) As the disorder strength increases the spins retain more information about their initial state, indicating a transition towards MBL. (f) shows the dynamics of $\langle\sigma_{i}^{z}\rangle$ for the strongest applied disorder, $W=8J_{max}$. The initial and steady-state time-averaged reduced density matrices for ions 1 and 10 now show that information is preserved about the initial spin configuration at the end of the evolution. Statistical error bars (1 s.d.) are smaller than the data points.}

\label{fig:Measuredevidenceoflocalization}
\end{figure*}
}

\newcommand{\FigureThree}{
\begin{figure}
\centering
\includegraphics[width=\columnwidth]{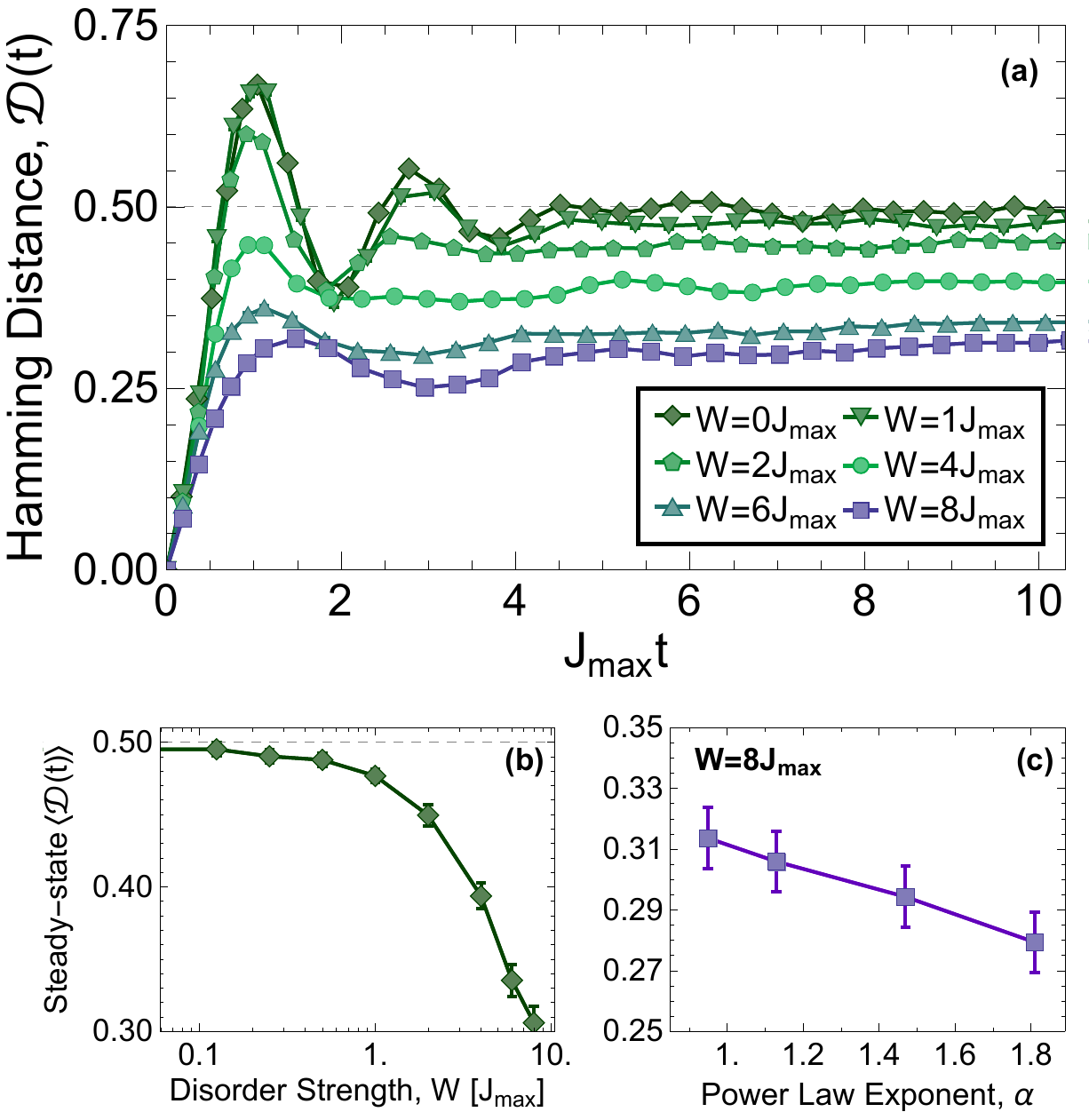}
\caption{\textbf{Hamming distance (HD).} (a) The Hamming Distance (HD) exhibits time dynamics that reach their steady-state values after $J_{max}t \approx 5$. For increasing disorder, the system becomes more strongly localized, and the steady-state Hamming Distance decreases. (different colors represent different disorder strengths). (b) The steady-state HD with respect to the strength of the random potential indicates the state is not localized for small disorder, but after the random field is sufficiently strong it becomes more localized with increased disorder. (c) The system becomes less localized in the presence of longer-range interactions (smaller $\alpha$). Error bars, 1 s.d.  }

\label{fig:Hammingdistance}
\end{figure}
}

\newcommand{\FigureFour}{
\begin{figure}
\centering
\includegraphics[width=\columnwidth]{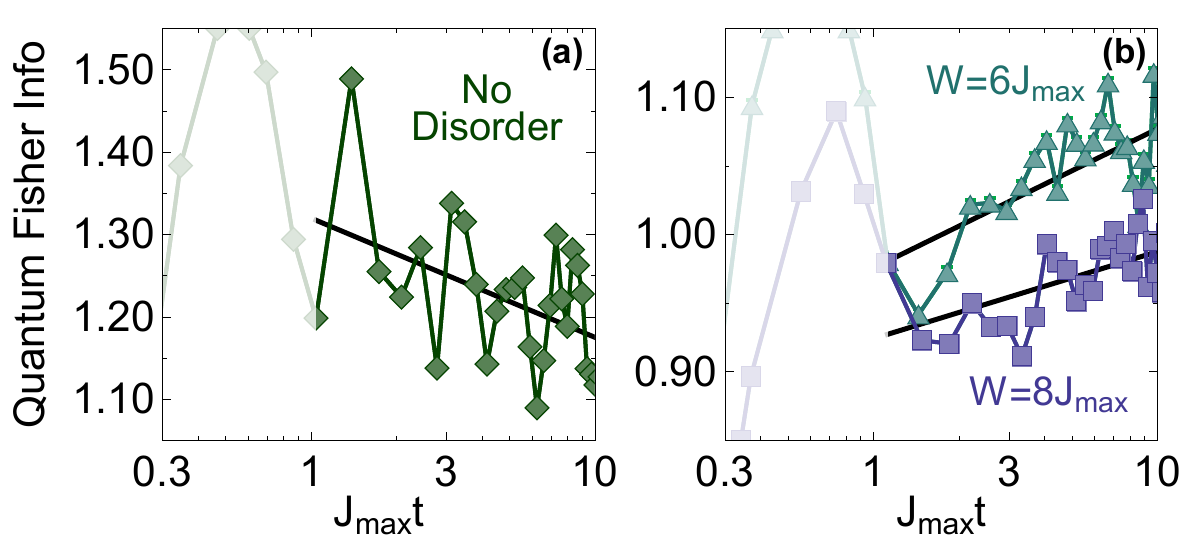}
\caption{\textbf{Quantum Fisher Information (QFI).} (a) The time evolution of the QFI for no disorder where there is no long-time growth of entanglement. The shaded area indicates the fast initial growth of QFI that follows a Lieb-Robinson-type bound. (b) The long-time logarithmic growth of the QFI for the applied disorder of $W=(6,8)J_{max}$ is a lower bound for the entanglement growth in the MBL state. Black lines are logarithmic fits to the data. Statistical error bars (1 s.d.) are smaller than the data points.}

\label{fig:QFI}
\end{figure}
}

\newcommand{\SuppFigureOne}{
\begin{figure*}
\centering
\includegraphics[scale=1]{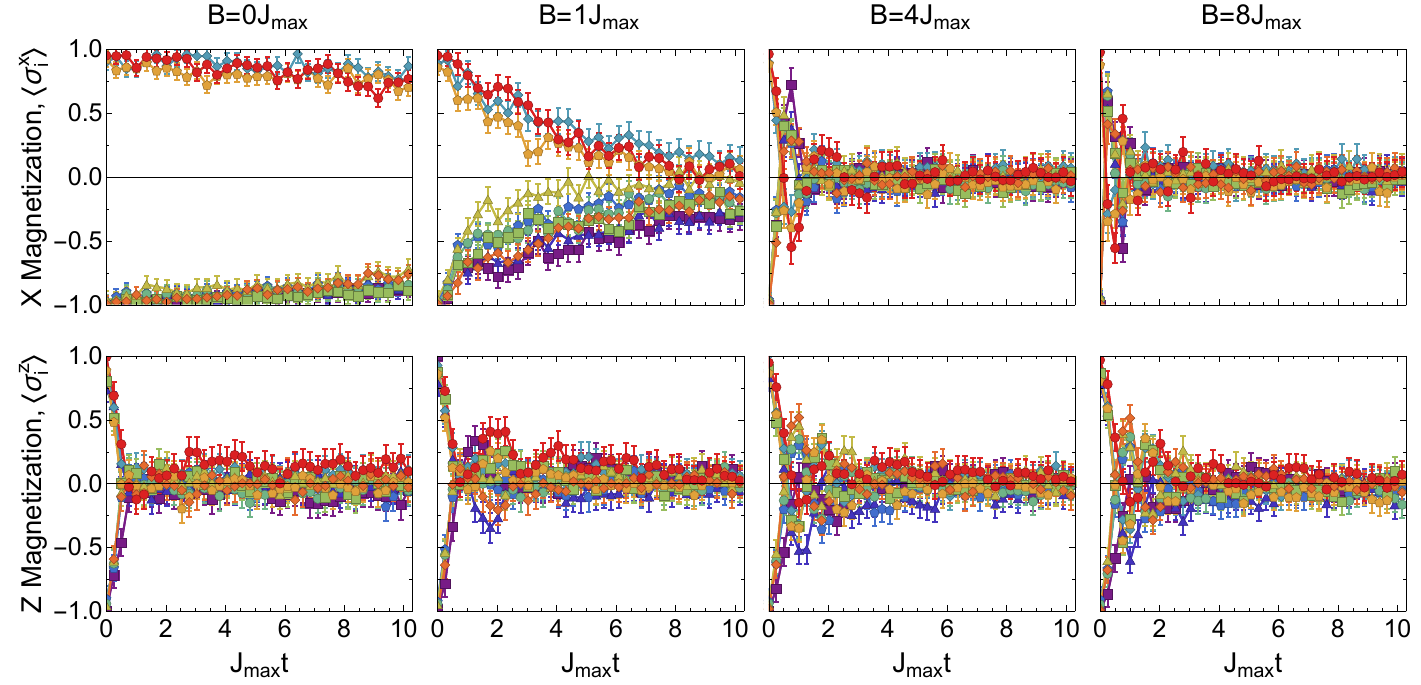}
\caption{\textbf{Measured thermalization in the transverse field Ising model.} The upper panels show the time dynamics of $\langle\sigma_i^x\rangle$ (different colors represent different ions) for 10 spins prepared in the random product state $\ket{\downarrow\downarrow\downarrow\uparrow\downarrow\downarrow\downarrow\uparrow\downarrow\uparrow}_x$, for different transverse magnetic field strengths. For $B=0$ the spins are in a eigenstate and do not thermalize. However, as the strength of $B$ is increased the system begins to thermalize more robustly and quickly. The lower panel plots the time evolution of $\langle\sigma_i^z\rangle$ with 10 spins prepared in the N\'eel ordered in the z direction for different transverse magnetic field strengths. We conclude that the system is in the thermalizing regime for $B=4J_{max}$ since we observe thermalizing behavior along two orthogonal directions. Error bars are 1 standard deviation of statistical error. }

\label{fig:Thermal}
\end{figure*}
}

\newcommand{\SuppFigureTwo}{
\begin{figure}
\centering
\includegraphics[width=\columnwidth]{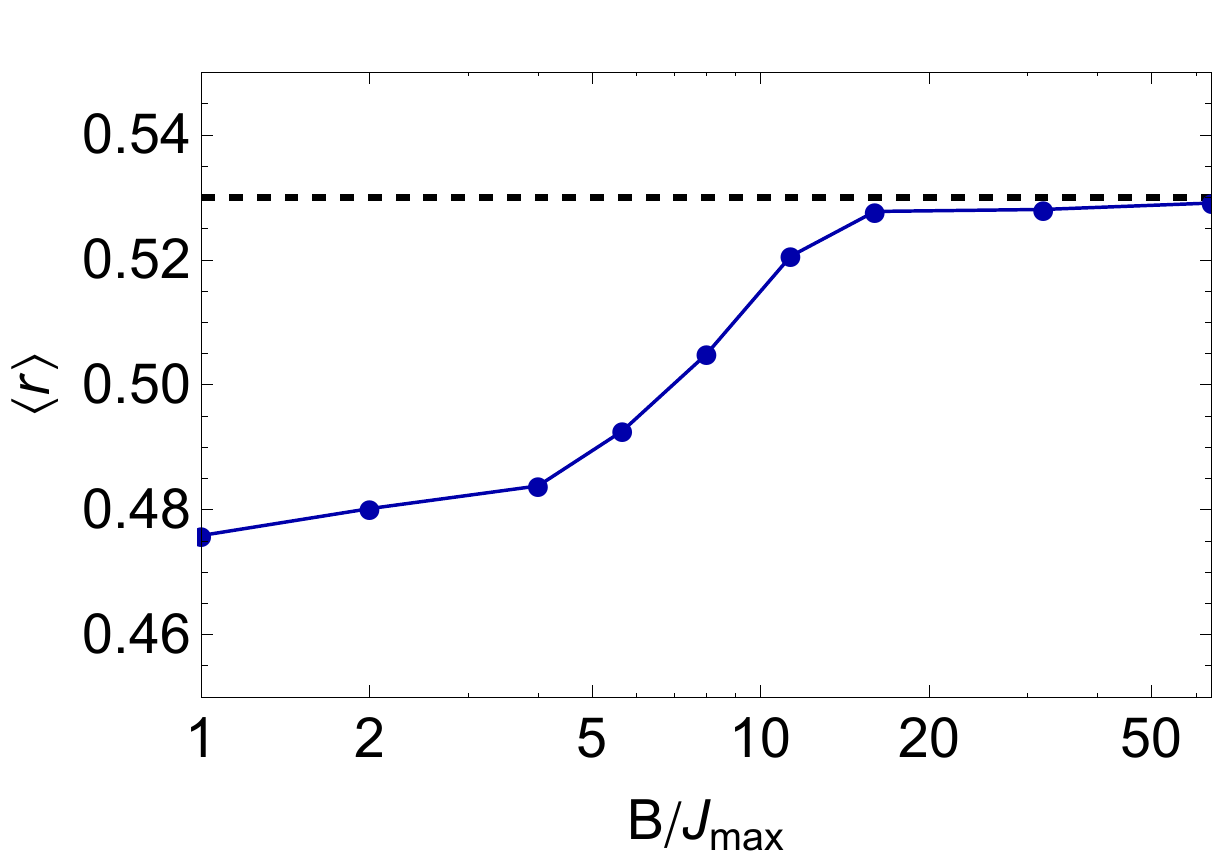}
\caption{\textbf{Thermalizing level statistics.} The calculated value of  $\langle r\rangle$ with respect to B saturates close to the predicted value for a random-matrix distribution (dashed black line) implying that the Hamiltonian is in the thermal phase for sufficiently large B.}

\label{fig:Level Statistics}
\end{figure}

}

\newcommand{\SuppFigureThree}{
\begin{figure}
\center{\includegraphics[width=\columnwidth]{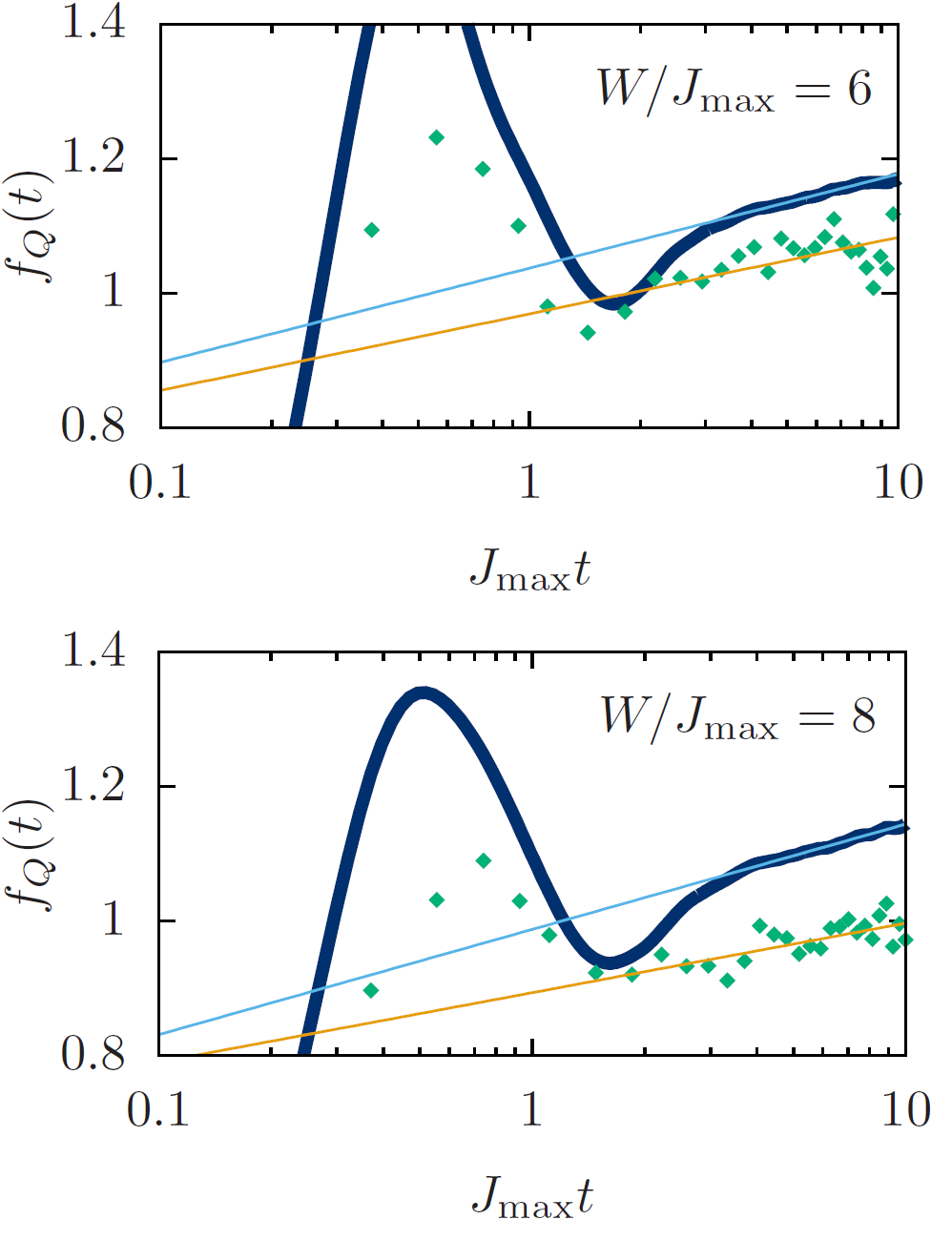}}
\caption{
\textbf{Comparison of the experimental data (green dots) with exact numerical simulations (thick blue lines) for QFI.} The solid straight lines represent logarithmic fits to the numerical (light blue) and experimental data (orange). Deviations from the ideal coherent dynamics due to decoherence in the experimental setup lead to a reduction of entanglement. Importantly, this suggests that the loss of purity does not generate a false positive for entanglement.
\label{fig:QFIexp}
}
\end{figure}

}

\newcommand{\SuppFigureFour}{
\begin{figure}
\center{\includegraphics[width=\columnwidth]{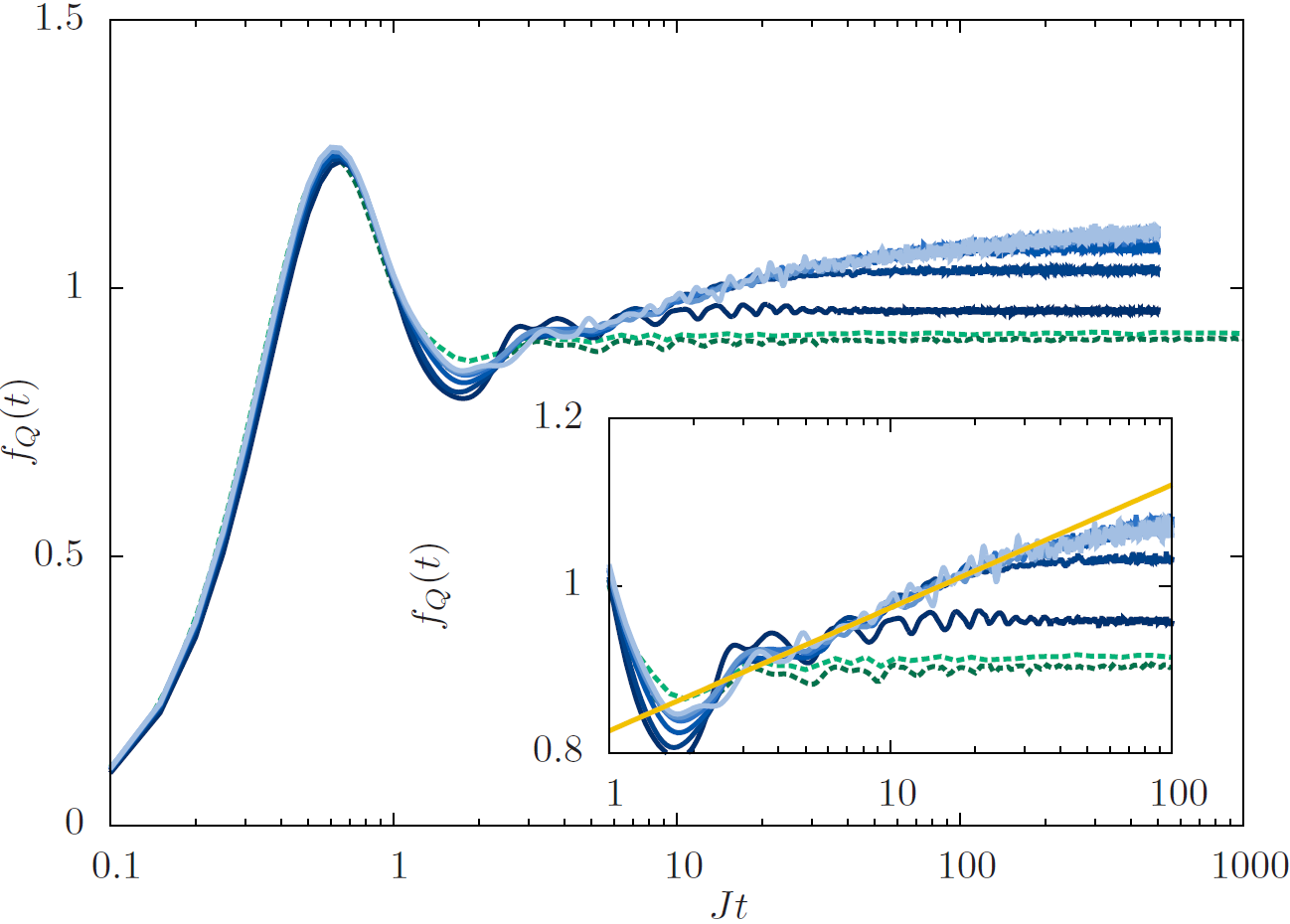}}
\caption{
\textbf{QFI from exact diagonalization ($\alpha=3$).} When subject to disorder, the QFI of the staggered magnetization shows a characteristic growth of entanglement (blue lines; from dark to light: $N=4,6,\dots,14$ averaged over $80000,20000,\dots,200$ disorder realizations). This growth is absent in a theory of free fermions with long-range hopping and pairing (green dashed lines with $N=14$ (dark green) averaged over $10000$ realizations and $N=100$ (light green) averaged over $1000$ realizations). Inset: In a time window dominated by next-nearest neighbor interactions, $2^\alpha<tJ<3^\alpha$, one observes a characteristic logarithmic entanglement growth, expected for a MBL system with short-range interactions.
\label{fig:QFIalpha3}
}
\end{figure}

}

\newcommand{\SuppFigureFive}{
\begin{figure}
\center{\includegraphics[width=\columnwidth]{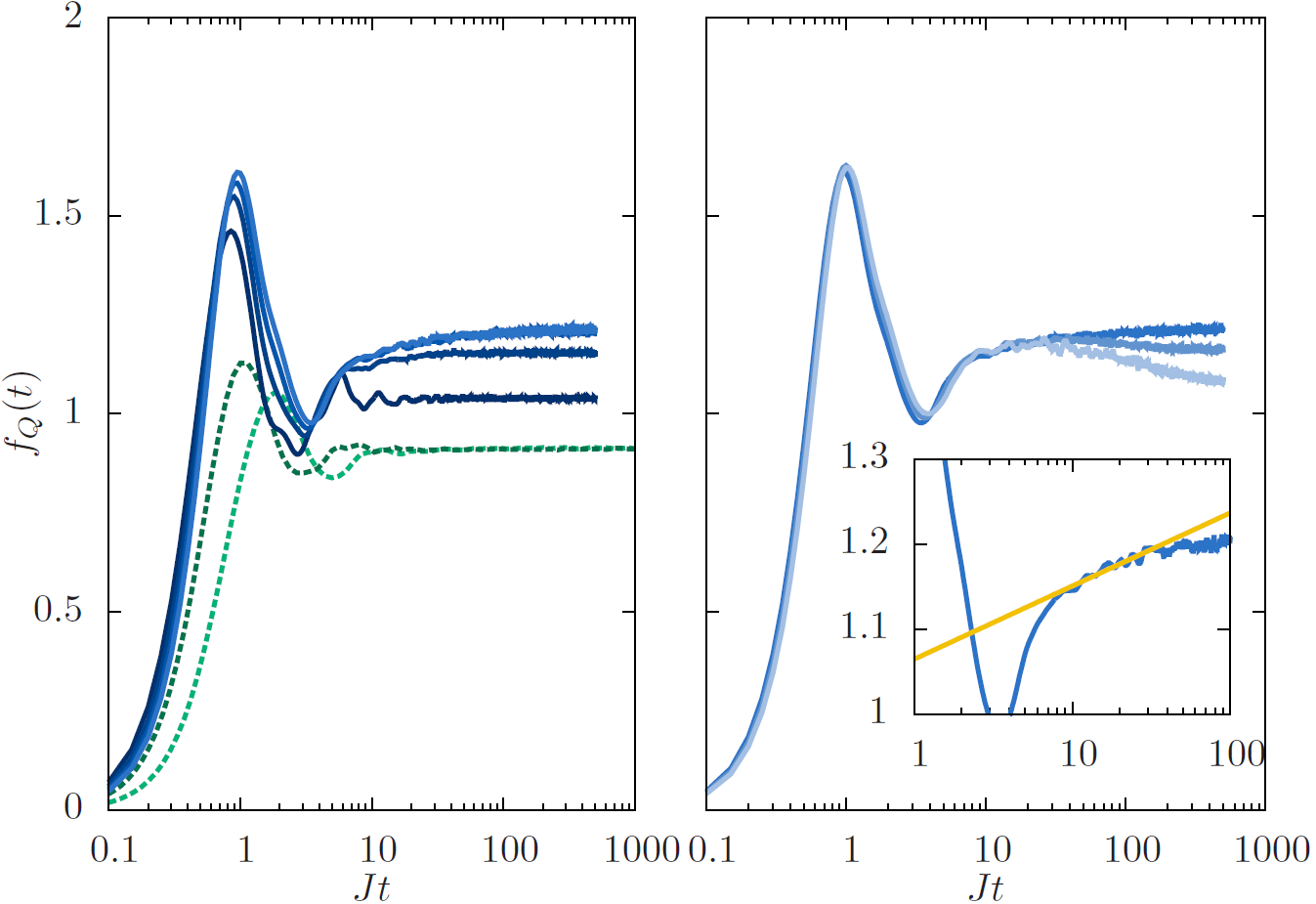}}
\caption{
\textbf{QFI from exact diagonalization ($\alpha=1.13$ and $W/J=3$).} Same color coding as in Fig.~\ref{fig:QFIalpha3}. (a) For small system sizes $N=4,\dots,10$, we observe the characteristic growth of entanglement as for $\alpha=3$ (compare with Fig.~\ref{fig:QFIalpha3}). (b) For slightly larger systems ($N>10$), a crossover appears at large times followed by a decrease of entanglement. This behavior might point towards a crossover to ergodicity in the thermodynamic limit. Importantly, however, for the experimentally relevant system size of $N=10$, we still find a time window consistent with a logarithmic growth of entanglement (see inset). Notice the difference of about a factor of $2$ in the time scale with respect to the main text.
\label{fig:QFIalpha1-13}
}
\end{figure}
}

\title{Many-body localization in a quantum simulator with programmable random disorder}

\author{J. Smith,$^{1}$ A. Lee}
\affiliation{Joint Quantum Institute, University of Maryland Department of Physics and National Institute of Standards and Technology, College Park, MD  20742}

\author{P. Richerme}
\affiliation{Department of Physics, Indiana University, Bloomington, IN, 47405}

\author{ B. Neyenhuis,$^{1}$ P. W. Hess,$^{1}$ P. Hauke,$^{3,4}$ M. Heyl}
\affiliation{Institute for Quantum Optics and Quantum Information of the Austrian Academy of Sciences, 6020 Innsbruck, Austria}
\affiliation{Institute for Theoretical Physics, University of Innsbruck, 6020 Innsbruck, Austria}

\author{D. Huse}
\affiliation{Physics Department, Princeton University, Princeton, NJ 08544, USA}

\author{C. Monroe}
\affiliation{Joint Quantum Institute, University of Maryland Department of Physics and National Institute of Standards and Technology, College Park, MD  20742}

\date{\today}

\begin{abstract}
When a system thermalizes it loses all local memory of its initial conditions. This is a general feature of open systems and is well described by equilibrium statistical mechanics.
Even within a closed (or reversible) quantum system, where unitary time evolution retains all information about its initial state, subsystems can still thermalize using the rest of the system as an effective heat bath. Exceptions to quantum thermalization have been predicted and observed, but typically require inherent symmetries \cite{Kinoshita2006, Gring2012} or noninteracting particles in the presence of static disorder \cite{Anderson1958,Wiersma1997, Billy2008, Roati2008}. The prediction of many-body localization (MBL), in which disordered quantum systems can fail to thermalize in spite of strong interactions and high excitation energy, was therefore surprising and has attracted considerable theoretical attention \cite{Anderson1958,Basko2007,Oganesyan2007, Pal2010, Serbyn2013}.
Here we experimentally generate MBL states by applying an Ising Hamiltonian with long-range interactions and programmably random disorder to ten spins initialized far from equilibrium. We observe the essential signatures of MBL:  memory retention of the initial state, a Poissonian distribution of energy level spacings, and entanglement growth in the system at long times. Our platform can be scaled to higher numbers of spins, where detailed modeling of MBL becomes impossible due to the complexity of representing such entangled quantum states.  Moreover, the high degree of control in our experiment may guide the use of MBL states as potential quantum memories in naturally disordered quantum systems \cite{Nandkishore2015}.
\end{abstract}

\maketitle

\FigureOne

It is exceedingly rare in nature for systems to localize, or retain local information about their initial conditions at long times. In an important counterexample, Anderson demonstrated that localization can arise due to the presence of disorder, which can destructively scatter propagating waves and prevent transport of energy or particles \cite{Anderson1958}. Although this interference effect can be applied to generic quantum systems, most experimental work has been restricted to the narrow parameter regime of low excitation energies and no interparticle interactions \cite{Wiersma1997, Billy2008, Roati2008}.
 
Whether such localization persists in the more general case of arbitrary excitation energy and non-zero interparticle interactions was theoretically explored by Anderson \cite{Anderson1958}, and more recently by others \cite{Basko2007,Oganesyan2007, Pal2010, Serbyn2013}. This MBL phase is predicted to emerge for a broad set of interaction ranges and disorder strengths, though the precise phase diagram is not well known \cite{Yao2014} since equilibrium statistical mechanics breaks down in the MBL phase and numerical simulations are limited to $\sim 20$ particles \cite{Oganesyan2007,Pal2010}. Very recent experiments searching for MBL have measured constrained mass transport and the breakdown of ergodicity in disordered atomic systems with interactions \cite{Kondov2015,Schreiber2015}.

\FigureTwo

Here we report the direct observation of MBL in a long-range transverse field Ising model with programmable, random disorder. This is a non-integrable model that cannot be mapped to noninteracting particles (a necessary condition for MBL \cite{Basko2007}) and we can easily tune the disorder strength and interaction range over a parameter space that exhibits this phenomenon. Our experiment is effectively a closed quantum system over the timescales of interest, since the system localizes approximately ten times faster than the coupling rate to the outside environment.

Each of the effective spin-1/2 particles is encoded in the $^2$S$_{1/2}\ket{F=0,m_F=0}$ and $\ket{F=1,m_F=0}$ hyperfine `clock' states of a $^{171}$Yb$^+$ ion, denoted $\ket{\downarrow}_z$ and $\ket{\uparrow}_z$, respectively \cite{YbDetection}.  We confine a chain of $10$ ions in a linear rf Paul trap and apply optical dipole forces to generate the effective spin-spin coupling \cite{Molmer1999} of a disordered Ising Hamiltonian:

\begin{equation}
\label{eqn:Hamiltonian}
H_\text{Ising}=\sum_{i<j} J_{i,j}\sigma_i^x\sigma_j^x+\frac{B}{2}\sum_{i}\sigma_i^z+\sum_{i}\frac{D_i}{2}\sigma_i^z
\end{equation}
where $\sigma_i^\gamma$ ($\gamma=x,z$) are the Pauli matrices acting on the $i^\text{th}$ spin, $J_{i,j}$ is the coupling strength between spins $i$ and $j$, $B$ is a uniform effective transverse field, $D_i$ is a site-dependent disordered potential, and $\hbar=1$ (Methods). After the chain evolves for some time, we collect the state-dependent fluorescence on an intensified charge-coupled device camera for site-resolved imaging. This, in addition to our ability to perform high fidelity rotations, allows measurement of the single-site magnetization $\langle\sigma_i^\gamma\rangle$ ($\gamma=x,y,z$) as well as arbitrary spin correlation functions along any direction. 

$J_{i,j}$ is a tunable, long-range coupling that falls off approximately algebraically as $J_{i,j} \propto J_{max}/|i-j|^\alpha$ \cite{QSIM2013Science}, where $J_{max}$ is typically $2\pi (0.5$ kHz$)$. Here we tune $\alpha$ between 0.95 and 1.81, although for most of the data $\alpha\approx1.13$. Moreover, we directly measure the complete spin-spin coupling matrix (Fig.\ \ref{fig:Interacting}a), demonstrating the long-range interactions required to exhibit MBL.

The site-specific programmable disorder term $D_i$ is sampled from a uniform random distribution with $D_i\in[-W,W]$. The disorder is generated by site-dependent laser-induced Stark shifts (Methods), which also allow for preparation of the system into any desired product state. To ensure we observe the general behavior of the disordered Hamiltonian, we average over 30 distinct random instances of disorder (Fig.\ \ref{fig:Interacting}b-c), which leads to a sampling error that is smaller than the features of interest in the data.

An important signature of the MBL phase is manifested in the spectral statistics of adjacent energy levels of the Hamiltonian. In a thermalizing phase, these energy splittings follow random-matrix level statistics due to level repulsion. However, in the MBL phase, this level repulsion is greatly suppressed since eigenstates typically differ by multiple spins flips. As a result, the level spacing between adjacent energy eigenvalues are Poisson-distributed \cite{Oganesyan2007,Pal2010}.  Using our directly measured spin-spin couplings and applied realizations for the strongest experimental disorder $W=8J_{max}$ and $B=4J_{max}$, we calculate the distribution of adjacent energy level splittings and find them to be Poisson-distributed, as expected for a MBL state (Fig.\ \ref{fig:Interacting}d). 

Before searching for evidence of localization in the system's time evolution, we first find parameters that cause the measured state to thermalize in the absence of disorder. We increase the transverse field $B$ and look for conditions that result in the single-site magnetization along two orthogonal directions approaching and remaining at their thermal equilibrium values (Methods).

Figure\ \ref{fig:Measuredevidenceoflocalization}a shows the measured dynamics of $\langle\sigma_i^z\rangle$ for $B=4J_{max}$ and $D_i=0$ with the spins initialized in the N\'eel ordered state, $\ket{\uparrow\downarrow\uparrow\downarrow\uparrow\downarrow\uparrow\downarrow\uparrow\downarrow}_z$  along the $z$-direction. This configuration has an energy equivalent to an infinite temperature thermal state, since the expectation value of the Hamiltonian is zero. At long times, each expectation value $\sigma_i^z$ approaches zero, losing memory of the initial ordering. As the transverse field $B$ is increased, the system appears to thermalize more quickly and the level statistics approach those of random matrices rather than Poissonians, as expected for a generic thermodynamic system (Methods).

\FigureThree

When $B\gg J$, the Hamiltonian is effectively an XY model \cite{Richerme2014,Jurcevic2014} and conserves $\sum_{i}\sigma_i^z$, because Ising processes that flip spins along the large field are energetically forbidden. Thus, being in a spin configuration with half of the spins up and half of the spins down maximizes the accessible energy states. In addition, the Ne\'el state is never an eigenstate, even for $B\gg J$ and $W\gg J$, since the uniform $B$ field at each site still allows spin exchange in the \textit{z}-basis. 

If a system is thermal, the Eigenstate Thermalization Hypothesis (ETH) provides a general framework where observables reach the value predicted by the microcanonical ensemble \cite{Deutsch1991, Srednicki1994, Rigol2008}. This allows us to calculate the expected thermal value of the reduced density matrix given the Hamiltonian and an initial state (Methods). To further establish that the system is thermalizing, we measure the reduced density matrix for each spin, $\rho_i=$Tr$_{\{j \neq i\}} \rho$, without applied disorder and $B=4J_{max}$ as shown in Fig.\ \ref{fig:Measuredevidenceoflocalization}a.  In our experiment, the spins are initially prepared in a product state with high fidelity. However at long times, the measured reduced density matrices show that each of the spins are very close to the zero magnetization mixed state, implying the system has locally thermalized. 

We apply the random disordered potential, $D_i \neq0$, and observe the emergence of MBL as we increase the strength of disorder. Since the many-body eigenstates in the MBL phase are not thermal, transport of energy and spins is suppressed, and ETH fails. Thus, observables will not relax to their thermal values \cite{Pal2010} and there will be memory of the initial conditions evident in the single-site magnetization. When starting in the Ne\'el ordered state, Fig.\ \ref{fig:Measuredevidenceoflocalization}b-f shows the time evolution of $\sigma_i^z$ for different disorder strengths. The frozen moments of the spins increase with increasing disorder as the emergent integrals of motion become more strongly localized \cite{Serbyn2013}.

With the maximum applied disorder, $W=8J_{max}$, we measure the single-spin reduced density matrix for the initial state and the averaged matrix for $J_{max}t\geq5$. In this case, localization of the spins leads to a marked difference in the measured and thermal reduced density matrices, indicating memory of the system's initial conditions and a breakdown of ETH. 

To quantify the localization, we measure the normalized Hamming distance (HD) \cite{Hauke2014}:

\begin{equation}
\label{eqn:Hilbert Space Distance}
\mathcal{D}(t)=\frac{1}{2}-\frac{1}{2N}\sum_i\bra{\psi_0}\sigma_{i}^z(t)\sigma_{i}^z(0)\ket{\psi_0}
\end{equation}
which gives the number of spin flips away from the initial state, normalized by the length of the chain. At long times, the HD approaches 0.5 for a thermalizing state and remains at 0 for a fully localized state. In Fig.\ \ref{fig:Hammingdistance}a, we measure that the long-time HD is 0.5 in the absence of disorder, and becomes smaller as the disorder strength is increased and the system more strongly localizes.

Figure\ \ref{fig:Hammingdistance}b shows that for finite but weak disorder, the time-averaged HD for $J_{max}t>5$ is unchanged, indicating no localization.  However, once the random field is sufficiently strong we observe a crossover from a thermalizing to a localized state. Once in this regime, the system becomes more localized with increasing disorder strength. 

There is great theoretical interest in mapping the MBL phase diagram with respect to interaction range and disorder strength \cite{Yao2014, Hauke2014, Burin2015}. We have taken the first steps towards this goal by measuring a change in the time-averaged HD for $W=8J_{max}$ and $J_{max}t>5$ as we adjusted the interaction range, $0.95<\alpha<1.81$ (Fig.\ \ref{fig:Hammingdistance}c). For shorter-range interactions, the system appears more localized, since the state approaches a fully-localized Anderson insulator as $\alpha \rightarrow \infty$. 

Although there are predictions of a many-body delocalization transition at $\alpha=1.5$ for Hamiltonians similar to ours, we did not observe this effect as we tuned $\alpha$ across this boundary. The lack of a sharp transition, along with the presence of MBL states for $\alpha < 1.5$, may be due to finite size effects (Methods). As this system is scaled to many dozens of spins, it will allow better study of the phase transition and mapping of the phase boundary in a regime where numerics are intractable.

A hallmark of MBL is the characteristic growth of entanglement under coherent time evolution \cite{Nandkishore2015}, though its experimental observation has been elusive so far. In Anderson insulators without many-body interactions, the entanglement production from weakly entangled initial states shows a quick saturation after a sharp transient regime. However, in MBL systems a long-time growth sets in, which is logarithmically slow for short-range interactions \cite{Bardarson2012} and can become algebraic with power-law interactions \cite{Pino2014}. 

This entanglement growth can be measured using a suitable witness operator or even full state tomography \cite{Haeffner2005}. We instead characterize the entanglement growth in this system by measuring the quantum Fisher information (QFI) \cite{Helstrom1976,Braunstein1994,Pezze2014}. The QFI gives a lower bound on the entanglement in the system while only requiring a measurement of two-body correlators, which can be efficiently accessed with our site-resolved imaging. Importantly, the QFI is able to  distinguish MBL from single particle localization via the anticipated characteristic entanglement growth (Methods). With no applied disorder, we observe a fast initial growth of the QFI following a Lieb-Robinson bound \cite{Richerme2014,Jurcevic2014} as the correlations propagate through the system, but no further growth afterwards (Fig.\ \ref{fig:QFI}a). In contrast, for the cases of applied disorder of $W = 6J_{max}$ and $W = 8J_{max}$, the further growth of the QFI is consistent with a logarithmic increase of entanglement at long times in a MBL state (Fig.\ \ref{fig:QFI}b), but is absent for single particle localized systems.  

\FigureFour

We have presented the experimental observation of MBL states in a quantum simulator with long-range interactions and random disorder. In a system whose level statistics predict MBL behavior, we observe a crossover between thermal and localized regimes as we increase the strength of applied random disorder, and we witness a long-time growth of entanglement in the localized state.  Our experimental platform is well suited for studying deep and intractable questions about thermalization and localization in quantum many-body systems.

\section{Acknowledgements}
We thank Luming Duan, Zhe-Xuan Gong, Tarun Grover, Chris Laumann, Shengtao Wang, Norman Yao, and Jiehang Zhang for helpful discussions. This work is supported by the ARO Atomic and Molecular Physics Program, the AFOSR MURI on Quantum Measurement and Verification, the IARPA MQCO program, and the NSF Physics Frontier Center at JQI. PH and MH acknowledge  the Deutsche Akademie der Naturforscher Leopoldina (grant No.~LPDS 2013-07), the EU IP SIQS, the SFB FoQuS (FWF Project No.~F4016-N23) and the ERC synergy grant UQUAM.

\bibliography{qsimrefs}

\section{METHODS}
\subsection{Generating the effective Hamiltonian}
We generate spin-spin interactions by applying spin-dependent optical dipole forces to ions confined in a 3-layer linear Paul trap with a 4.8 MHz radial frequency. Two off-resonant laser beams with a wavevector difference $\Delta \vec{k}$ along a principal axis of transverse motion globally address the ions and drive stimulated Raman transitions. The two beams contain a pair of beatnote frequencies symmetrically detuned from the resonant transition at $\nu_0=12.642819$ GHz by a frequency $\mu$, comparable to the transverse motional mode frequencies. In the Lamb-Dicke regime, this results in the Ising-type Hamiltonian in Eq. (1) \cite{Molmer1999,PorrasCiracQSIM,QSIMPRL2009} with
\begin{equation}
\label{eqn:Jij}
J_{i,j}=\Omega^2\omega_R \sum_{m=1}^N \frac{b_{i,m}b_{j,m}}{\mu^2-\omega_m^2},
\end{equation}
where $\Omega$ is the global Rabi frequency, \mbox{$\omega_R=\hbar\Delta k^2/(2M)$} is the recoil frequency, $b_{i,m}$ is the normal-mode matrix \cite{James1998}, and $\omega_m$ are the transverse mode frequencies. The coupling profile may be approximated as a power-law decay $J_{i,j}\approx J_{max}/|i-j|^\alpha$, where in principle $\alpha$ can be tuned between 0 and 3 by varying the laser detuning $\mu$ or the trap frequencies $\omega_m$ \cite{PorrasCiracQSIM,QSIM2013Science}. In this work, $\alpha$ was tuned approximately between 0.95 and 1.81 by changing $\mu$. By asymmetrically adjusting the laser beatnote detuning $\mu$ about the carrier by a value of $B$ we apply a global Stark shift that can be thought of as a uniform effective transverse magnetic field of $({B}/{2})\sigma_{i}^{z}$ .

\SuppFigureOne

We generate the effective disorder by applying a site-dependent Stark shift using a single 355nm laser beam that is focused down tightly to a $1/e^2$ waist of $\sim 1.8\mu m$. The ion separation is $\sim 2.5 \mu m$, thus the crosstalk between ions is negligible with a measured ratio of nearest-neighbor Rabi frequencies of $\sim20:1$. We use an acousto optic modulator (AOM)  with a full width at half maximum bandwidth of $\approx$100 MHz to apply the Stark shift to each ion. The AOM is not imaged onto the ions, so that driving the AOM with different frequencies allows the position of the beam to be scanned over the length of a 10 ion chain, $\sim 20\mu m$. The Stark shift is proportional to $I^2$. Thus, to achieve larger applied Stark shifts, we raster through the AOM drive frequencies corresponding to addressing each ion with a total cycle time of $\sim5 \mu s$ instead of applying them simultaneously.  Since
we cannot control the sign of the site-specific Stark shift, to center the disorder strength around the global transverse field, we adjust the asymmetric detuning by $WJ_{max}/2$.

\subsection{Measuring the spin-spin coupling matrix}
In order to observe the dynamics between just two of the ions in the chain, we shelve the other spins out of the interaction space. This is done by performing a $\pi$ rotation between $\ket{\downarrow}_z$, $^2$S$_{1/2}\ket{F=0,m_F=0}$, and one of the Zeeman states, $^2$S$_{1/2}\ket{F=1,m_F=-1}$, while shifting the two ions of interest out of resonance by applying a large Stark shift with the individual addressing beam. We then apply our Hamiltonian which now acts only on the two ions left in the interaction space and determine the elements of the spin-spin coupling matrix by fitting the measured interaction Rabi flopping frequency between each pair of spins.


\subsection{Arbitrary product state preparation}
 State initialization starts with optically pumping the spins with high-fidelity to  $\ket{\downarrow\downarrow\downarrow\cdots}_z$. Then we perform a global ${\pi}/{2}$ rotation to bring the ions to $\ket{\downarrow\downarrow\downarrow\cdots}_x$. At this point we apply a Stark shift with the individual addressing beam to the spins that are to be flipped and allow the chain to evolve until these ions are $\pi$ out of phase with rest of the ions. This, along with our ability to perform high fidelity global rotations, allows for the preparation of any arbitrary product state along any direction of the Bloch sphere. Individual spin flips can be achieved with a fidelity of $\sim0.97$, while arbitrary state preparation can be done with a fidelity of $\sim(0.97)^N$, where N is the number of spins flipped with the individual addressing beam.

\subsection{Determining a Set of Thermalizing Parameters} 
Extended Data Figure\ \ref{fig:Thermal} shows the time evolution of $\langle\sigma_{i}^{x}\rangle$ for different values of B for the spins initialized in the randomly chosen product state $\ket{\downarrow\downarrow\downarrow\uparrow\downarrow\downarrow\downarrow\uparrow\downarrow\uparrow}_x$. Without a transverse field, the spins are in an eigenstate of the Ising interaction and undergo no evolution. Once a transverse field is added the individual spins begin to lose memory of their initial conditions and as its strength is increased, the ions thermalize faster and more robustly. 

To confirm the system is thermalizing, we measure the time evolution of the single site magnetization, $\langle\sigma_{i}^{z}\rangle$, along an orthogonal direction for different strengths of the transverse magnetic field starting with the spins initialized in the N\'eel ordered state. As seen in Extended Data Fig.\ \ref{fig:Thermal} the spins have lost information about their initial conditions in the \textit{z} direction for all values of B.

We calculate the spectral statistics of adjacent energy levels for the Hamiltonian and find they are not Poisson distributed for $B=4J_{max}$ and $D_i=0$ indicating that with no applied disorder, the system is not in a localized phase. Furthermore, one can determine if a system is in a thermal or localized regime by finding the correlation between adjacent energy splittings by calculating the ratio of two consecutive gaps \cite{Oganesyan2007}:
\begin{equation}
r_{n}=\frac{min\{\delta_n,\delta_{n-1}\}}{max\{\delta_n,\delta_{n-1}\}}
\end{equation} 
where $\delta_n=E_{n+1}-E_{n}\geq0$. For a localized phase, where one expects a Poisson energy spectrum, the probability distribution of this order parameter is given by $P_p(r)=2/(1+r)^2$ and thus $\langle r\rangle \approx0.39$. For energy level spacings following a random-matrix as predicted for a thermalizing regime, we calculate $\langle r\rangle\approx0.53$ for a chain of 10 spins. Extended Data Figure \ref{fig:Level Statistics} shows that $\langle r\rangle$ saturates to the expected value for a random matrix distribution, indicating that the Hamiltonian is thermal for sufficiently large B.

\SuppFigureTwo

\subsection{Calculating the density matrix expected by the Eigenstate Thermalization Hypothesis}
Given a Hamiltonian and an initial state $\ket{\psi_{0}}$, the corresponding energy is $\bra{\psi_{0}}H\ket{\psi_{0}}$. For a thermalizing system satisfying ETH this energy should be equal to the classical energy:
\begin{equation}
\label{eqn:Classical Energy}
E=\frac{\mathrm{Tr}[He^{-\beta H}]}{\mathrm{Tr}[e^{-\beta H}]}
\end{equation}
for the appropriate $\beta=1/(k_{B}T)$. When partitioning the entire system into subsystems A and B, with the size of A much smaller than B (perhaps even a single spin), then, the density matrix on site A at long times can be approximated by:
\begin{equation}
\label{eqn:Density Matrix}
\rho_A=\frac{\mathrm{Tr_{B}}[e^{-\beta H}]}{\mathrm{Tr}[e^{-\beta H}]}
\end{equation}

\SuppFigureThree

Since we start in the N\'eel ordered state, the initial energy given the Hamiltonian in Eq. (1) is equal to zero, $\bra{\psi_{0}}H\ket{\psi_{0}}=0$. Equating this to the right hand side of Eq.~\eqref{eqn:Classical Energy} and solving for $\beta$ gives $\beta=0$, or $T=\infty$. Using this $\beta$ in Eq.~\eqref{eqn:Density Matrix} gives a value for any reduced thermal density matrix of:
\[ \left( \begin{array}{cc}
1/2 & 0  \\
0 & 1/2 \\
\end{array} \right)\] 
in agreement with the measured reduced density matrices in Fig. 2a.

\subsection{Quantum Fisher Information}
The quantum Fisher information (QFI) has recently been shown to witness genuinely multipartite entanglement \cite{Hyllus2012,Toth2012}. From a quantum metrology perspective, the QFI quantifies the sensitivity of a given input state to a unitary transformation $e^{i\vartheta\obs}$ generated by the hermitian operator $\obs$. In a pure state, it is given by \cite{Braunstein1994}
\begin{equation}
\label{eq:FQ}
F_Q=4(\Delta\obs)^2 = 4 (\langle\obs^2\rangle-\langle\obs\rangle^2).
\end{equation}

For a local operator $\obs=\sum_{i=1}^N \obs_i$ (where the difference between largest and smallest eigenvalue of $\obs_i$ is $1$), the QFI witnesses entanglement as soon as 
\begin{equation}
f_Q\equiv F_Q/N>1\,.
\end{equation}

To characterize the growth of entanglement out of the initial N\'eel state, the natural choice of the generator $\obs$ is the staggered magnetization, $\obs=\sum_{i=1}^N (-1)^i \sigma_i^z/2$. Remarkably, this QFI is proportional to the variance of the Hamming distance ${\cal D}(t)$ given by Eq. (2) of the main text, 
\begin{equation}
\label{eq:FQstaggeredMagnetization}
F_Q=4N^2 (\Delta \hat{\cal D})^2=\sum_{i,j} [ (-1)^{i+j} \langle\sigma_i^z \sigma_j^z\rangle ] -[\sum_i(-1)^i \langle\sigma_i^z\rangle ]^2\,,
\end{equation}
when associating ${\cal D}(t)=\langle\hat{\cal D}(t)\rangle$, with $\hat{\cal D}= 1/(2N) [1-\sum_{i=1}^N (-1)^i \sigma_i^z ]$.

The QFI as defined in Eq.~\eqref{eq:FQ} assumes a pure state, i.e., that time evolution is purely unitary. For mixed states, the QFI cannot be expressed as a simple expectation value of the operator $\obs$ \cite{Braunstein1994}. In general, decoherence reduces the purity of the system's state over experimental time scales. To show that the measured increase of $F_Q$ as defined in Eq.~\eqref{eq:FQstaggeredMagnetization} is indeed due to coherent dynamics, we compare to numerically exact calculations for a unitary time evolution using the experimental parameters. As can be seen in Fig.~\ref{fig:QFIexp}, the experimental data is always below the theoretical prediction for a unitary time evolution. The loss of purity, therefore, does not generate a false positive indicator of entanglement in our system. Moreover, in the time window following the strong initial rise, the increase of $F_Q$ for both theory and experiment is consistent with a logarithmic growth (see below). 

\SuppFigureFour

To study how the localization behavior changes with system size, we performed a numerical finite-size scaling. In order to obtain a well-behaved scaling, we use the Kac prescription \cite{Kac1966}, i.e., we adjust the couplings as $J_{ij}=J  {\cal N}^{-1} \left|i-j\right|^{-\alpha}$, where 
${\cal N}=\left(N-1\right)^{-1}\sum_{i<j} \left|i-j\right|^{-\alpha}$. 
Note that using this prescription the fundamental energy scale $J$ differs by about a factor of $2$ from $J_{max}$, the value used in the main text.  

For $\alpha>2$, the disordered long-range Ising Hamiltonian shows MBL behavior at large disorder \cite{Burin2006}. In Fig.~\ref{fig:QFIalpha3}, we plot the dynamics of the QFI for $\alpha=3$, where it can be seen that the system displays a characteristic long-time growth of entanglement. In particular, within a time window $2^\alpha<tJ<3^\alpha$ where only next-nearest-neighbor interactions are relevant and next-to-next-nearest-neighbor interactions are still negligible, one expects the system to behave essentially as a nearest-neighbor Ising model with a weak next-to-nearest-neighbor coupling. For such a system, a logarithmic growth of entanglement is expected, as we indeed find in that regime, see inset in Fig.~\ref{fig:QFIalpha3}. Moreover, in Fig.~\ref{fig:QFIalpha3}, we compare our numerical results for the spin chain to the appropriate long-range free fermionic theory (see below), which shows a quick system-size independent saturation of entanglement without a further growth. Therefore, we conclude that the observed growth of the QFI is not possible in a quantum system without many-body interactions, thus giving a clear signature for true MBL behavior.

\SuppFigureFive

The situation is more complex at $\alpha=1.13$. For the case $B=0$, it has been predicted that within the range $1<\alpha<2$ delocalized behavior could be expected in the thermodynamic limit \cite{Burin2006}. As seen in Fig.~\ref{fig:QFIalpha1-13}, for system sizes $N<12$ the model displays all essential signatures of MBL, as found for $\alpha=3$. Performing finite-size scaling, however, indicates a crossover to potentially ergodic behavior for larger chains. The growth of the QFI is cut off for larger $N$ followed by a decrease of the entanglement on longer time scales. The time scale for the crossover becomes smaller for increasing system size, which could in the thermodynamic limit potentially remove the characteristic entanglement growth completely. Hence, the system might become ergodic, but, unfortunately, the exact diagonalization quickly reaches its limits at this point. Here, scaling our quantum simulator to larger system sizes could thus resolve a difficult open question, namely of the existence of ergodicity in the range $1<\alpha<2$. However, we would like to emphasize that the essential features of MBL phases in long-range disordered Ising chains are nevertheless captured by the considered experimental system at $N=10$. In particular, we still find a time window consistent with a logarithmic growth of entanglement, see inset in Fig.~\ref{fig:QFIalpha1-13}.

To show that the entanglement growth is truly due to interactions, we also compare the exact data to a close approximation of $H$, Eq.~(1) of the main text, with a non-interacting theory. Using the Jordan-Wigner transformation, $\sigma_j^-\to e^{-i\theta_j} c_j$, with the phase of the string operator $\theta_j=\pi \sum_{j<i}c_j^\dagger c_j$, the Hamiltonian Eq.~(1) can be mapped to a fermionic theory with annihilation and creation operators $c_j$ and $c_j^\dagger$, respectively,
\begin{equation}
H=\sum_{i<j} J_{ij} (c_i^\dagger e^{i(\theta_j-\theta_i ) } c_j+c_i^\dagger e^{i(\theta_j+\theta_i )}  c_j^\dagger+h.c.)-\sum_i(B+D_i ) c_i^\dagger c_i \,.
\end{equation}
If $J_{ij}$ contained only nearest-neighbor interactions, this Hamiltonian would become equivalent to a free fermionic theory. For general $J_{ij}$, however, the string operators generate interactions between the fermions. Over short times, and especially in a localized regime, the phases $\theta_j$ are dominated by their initial values, i.e., it is a good approximation to replace (for the initial Neel state) $\theta_j\to \pi \sum_{j<i}((-1)^j+1)/2$ in the Hamiltonian. This replacement amounts to approximating $H$ by a non-interacting fermionic theory with long-range hopping and pairing. The QFI for that case is included in Figs.~\ref{fig:QFIalpha3} and~\ref{fig:QFIalpha1-13}. As one can see, the QFI quickly saturates to values below $f_Q=1$. The experimentally and numerically observed further growth of the QFI is thus truly due to interactions, and cannot be captured within a free fermionic theory, even with long-range hopping.

\end{document}